\documentclass{article}
\usepackage[preprint]{a2026}
\usepackage[utf8]{inputenc} 
\usepackage[T1]{fontenc}    
\usepackage{hyperref}       
\usepackage{url}            
\usepackage{booktabs}       
\usepackage{amsfonts}       
\usepackage{nicefrac}       
\usepackage{microtype}      
\usepackage{xcolor}         
\usepackage{amsmath}
\usepackage{graphicx}    
\usepackage{subcaption}  

\title{Dual-BEATs: Unlocking Zero-Shot Stereo Audio Perception in Audio Large Language Models via Dithering}
\setcitestyle{numbers,square}

\author{%
  Shuo-Chun Lin \\
  Institute of Information Science, Academia Sinica, Taiwan \\
  \texttt{shuochunlin@as.edu.tw} \\
  \And
  Hen-Hsen Huang \\
  Institute of Information Science, Academia Sinica, Taiwan \\
  \texttt{hhhuang@iis.sinica.edu.tw} \\
}

\begin{document}

\maketitle

\begin{abstract}
  Multimodal Large Language Models (LLMs) have remarkable semantic audio understanding, yet they remain ``spatially agnostic'' due to their reliance on mono-channel audio representations. 
  Currently, spatial audio perception methods mainly focus on complex room simulations and custom-trained, geometry-aware stereo encoders, which limits their accessibility and generalizability. 
  In this paper, we introduce the \textbf{Dual-BEATs} architecture, in which the left and right audio channels are routed independently through two identical semantic encoders as an alternative to specialized spatial modules. 
  To circumvent the architectural bottleneck where internal normalization otherwise erases the inter-channel variance of stereo audio, we inject a static, uncorrelated dithering noise floor prior to encoding. 
  This dithering intervention establishes a macro-variance floor that ``smuggles'' spatial geometry across the normalization layers. 
  Evaluated on a ternary directional classification task (Left, Center, Right), we demonstrate that dithered models achieve exceptional spatial resolution—reaching up to 97.2\% localization accuracy even on subtle 0.5 panning amplitudes—and demonstrates robust, zero-shot generalization to entirely unseen spatial configurations. 
  Our results suggest that with the appropriate acoustic regularization, standard multimodal models are natively capable of generalized stereo audio understanding.
\end{abstract}

\section{Introduction}
\label{introduction}

The rapid evolution of multimodal Large Language Models (LLMs) has fundamentally transformed computational audio perception. Foundational open-weight architectures such as Audio Flamingo~\citep{kong2024audio}, SALMONN~\citep{tang2023salmonn}, and established end-to-end systems such as Qwen2-Audio~\citep{chu2024qwen2} have demonstrated remarkable zero-shot capabilities across a wide range of audio-linguistic tasks. 
By mapping continuous audio waveforms to the discrete textual embedding space of an LLM, these models can successfully reason about complex speech, acoustic environments, and semantic events, exhibiting profound semantic depth.

However, despite these advanced semantic capabilities—exemplified by recent models such as Music Flamingo~\citep{ghosh2026music} which perform complex chain-of-thought reasoning over music theory—current foundational audio models operate under a restrictive architectural assumption: they process sound with no concept of physical space. The standard preprocessing pipeline for general audio LLMs routinely downmixes stereo or multi-channel audio into a single mono audio channel prior to encoding~\citep{radford2022whisper, elizalde2023clap}. 
This spatial deficit extends far beyond open-weight architectures—even state-of-the-art proprietary multimodal systems, such as the GPT-4 family~\citep{openai2023gpt4, openai2024gpt4o} and Gemini~\citep{gemini2023}, rely on preprocessing pipelines that inherently flatten audio into mono mel-spectrograms. Consequently, because these models are architecturally deaf to inter-channel geometry, any spatial localization they output is purely an artifact of semantic hallucination rather than true acoustic perception.

Recent efforts to address this spatial blindness often rely on complex 3D room simulations or custom-trained spatial encoders that are tightly coupled with their LLM backbones~\citep{peng2024spatialast, biswas2026owl}. 
While effective for specialized applications, these purpose-built architectures are highly sensitive to acoustic domain shifts. For instance, adapting an encoder optimized for meticulously simulated binaural environments to process standard, unsimulated stereo mixes can aggressively disrupt its pre-trained semantic alignment. A more intuitive and modular approach to generalized stereo perception bypasses these spatial encoders entirely by routing the left and right audio channels independently through two identical, standard semantic encoders, such as the BEATs architecture~\citep{chen2022beats}.

Yet, this straightforward dual-encoder approach remains largely unexplored. We hypothesize that this absence is due to a structural bottleneck within standard LLM architectures: internal normalization layers~\citep{ba2016layer, zhang2019root}. Because normalizers are designed to forcefully standardize the variance of incoming latent vectors, they inadvertently act as dynamic compressors when processing continuous audio. While absolute, hard-panned boundaries (Panning Amplitude, $PA = 0.0$) may present a large enough variance to survive this bottleneck, normalizers aggressively equalize the subtle inter-channel differences of intermediate panning amplitudes (e.g., $PA = 0.5$, a mere $\sim$6dB differential). This effectively erases continuous spatial geometry, rendering the model blind to subtle spatial gradients and destroying its ability to generalize.

Building on this modular approach, we propose the \textbf{Dual-BEATs} architecture augmented with a targeted acoustic intervention: a static, uncorrelated stereo dithering noise floor injected prior to encoding. This additive dither safely bridges the normalization layers, preserving the spatial amplitude delta by embedding it directly into the noise signature, thereby rendering the spatial geometry highly visible to the LLM's cross-attention mechanisms.

Through extensive empirical evaluation across distinct architectures and parameter scales (Gemma-3-1B and OLMo-3-7B) utilizing decoupled audio encoders, we demonstrate that this dithering intervention unlocks true spatial abstraction. Specifically, our primary contributions are twofold:

\begin{itemize}
\item \textbf{Zero-Shot Spatial Perception}: We demonstrate that standard multimodal models can achieve exceptional spatial resolution without geometry-aware encoders. Models regularized by the uncorrelated noise floor reach up to 97\% directional accuracy at subtle 0.50 panning amplitudes and exhibit robust, continuous zero-shot generalization to entirely unseen spatial configurations.
\item \textbf{Bypassing the Normalization Bottleneck}: We empirically identify the compressive limitations of internal normalization layers in continuous audio processing and establish uncorrelated dithering as a fundamental, signal-level intervention that preserves inter-channel geometry across distinct normalization topologies, circumventing the need for specialized spatial encoders.
\end{itemize}

\section{Related Works}
\label{literature}

\paragraph{General Audio LLMs}
Current audio-capable multimodal models typically rely on a standard tripartite architecture: a pre-trained acoustic encoder, a projection module, and the LLM backbone. Models such as Audio Flamingo~\citep{kong2024audio}, GAMA~\citep{ghosh2024gama}, and SALMONN~\citep{tang2023salmonn} demonstrate remarkable zero-shot reasoning capabilities. Recent advancements, such as Music Flamingo~\citep{ghosh2026music}, push these boundaries further to achieve near-human reasoning over complex acoustic structures. However, this semantic depth masks a fundamental limitation: general-purpose pipelines inherently assume a mono input. Multi-channel audio is routinely downmixed prior to encoding~\citep{radford2022whisper, elizalde2023clap, baevski2020wav2vec, hsu2021hubert}. Consequently, these bottlenecks render models structurally blind to spatial geometry; they can expertly reason about \textit{what} an acoustic event is, but not \textit{where} it is. Critiques like \textit{The World is Not Mono} \citep{you2026notmono} explicitly identify this downmixing as a critical barrier, arguing that genuine auditory scene analysis requires spatial intelligence.

\paragraph{Specialized Spatial Encoders}
To address this spatial deficit, a rapidly expanding ecosystem of methods has emerged across various input-output representations~\citep{zhu2025asaudio}, largely driven by downstream applications like the DCASE Sound Event Localization and Detection (SELD) challenges \citep{dcase2025seld}. Initial explorations \citep{chen2024can} demonstrated LLMs could achieve robust sound source localization given appropriate spatial cues. More recently, SAVVY \citep{chen2025savvy} introduced a training-free pipeline for 3D spatial reasoning in audio-visual contexts, though it relies on integrating external localization modules. This spurred the development of specialized acoustic encoders explicitly designed for complex ambisonics and geometry-aware topologies. Frameworks like Spatial-AST (BAT) \citep{peng2024spatialast} map spatial characteristics using interaural phase differences. To make these systems hardware-agnostic, PhaseCoder \citep{dementyev2026phasecoder} explicitly ingests microphone coordinate geometry alongside the audio stream. These dedicated architectures are increasingly deployed for reasoning tasks like spatial question answering and tracking dynamic source movements \citep{sridhar2025spatialmotion}. Recent hybrid architectures \citep{you2026notmono} extend this approach by deploying heterogeneous encoders in parallel—pairing a standard semantic model with a custom-built spatial encoder to explicitly combine meaning and geometry. While highly effective, these methods require significant architectural overhead, tightly coupled pre-training, or explicit geometric conditioning. A critical gap remains for generalized stereo perception directly from standard, off-the-shelf semantic encoders, without relying on heavy, phase-aware topologies.

\paragraph{The Normalization Constraint}
Processing left and right audio channels independently through identical semantic encoders offers an intuitive, decoupled path to stereo perception. However, applying this approach to modern LLMs presents a unique physical challenge. We hypothesize that standard internal regularization layers—such as variance-scaling normalizers \citep{ba2016layer, zhang2019root}—act as an aggressive bottleneck for continuous spatial audio. By forcefully standardizing the variance of incoming latent vectors, they inadvertently compress the microscopic inter-channel amplitude differentials required to perceive spatial panning. While noise injection is a well-established technique in deep learning—whether deployed to induce regularization during instruction tuning (e.g., NEFTune \citep{jain2024neftune}) or to study adversarial robustness \citep{carlini2017towards}—its application as a structural bypass mechanism for continuous multimodal signals remains largely unexplored. Drawing on the principles of Stochastic Resonance from classical physics and digital signal processing, where noise is utilized to elevate weak signals across functional thresholds \citep{benzi1981mechanism, vanderkooy1987dither}, we propose a novel architectural intervention. By injecting an uncorrelated dithered noise floor prior to encoding, we create a Stochastic Resonance bridge. This preserves the crucial left-right amplitude delta—the core geometric differential—across the normalization layers, unlocking spatiality without specialized encoders.

\section{Methodology}
\label{sec:methodology}

\subsection{Dual-BEATs Architecture and Spatial Routing}
Standard audio models typically downmix multi-channel waveforms into a single mono channel, destroying spatial geometry. To bypass this without relying on geometry-specific encoders, we propose a decoupled routing topology. Given a stereo input, we isolate the left and right channels ($W_L, W_R$). As formalized in Section 3.2, these signals are first augmented with uncorrelated dithering noise to form $W_L'$ and $W_R'$. We route these augmented signals through two identical, frozen BEATs encoders \citep{chen2022beats} individually. 

\begin{figure}[tbp]
    \includegraphics[width=\linewidth]{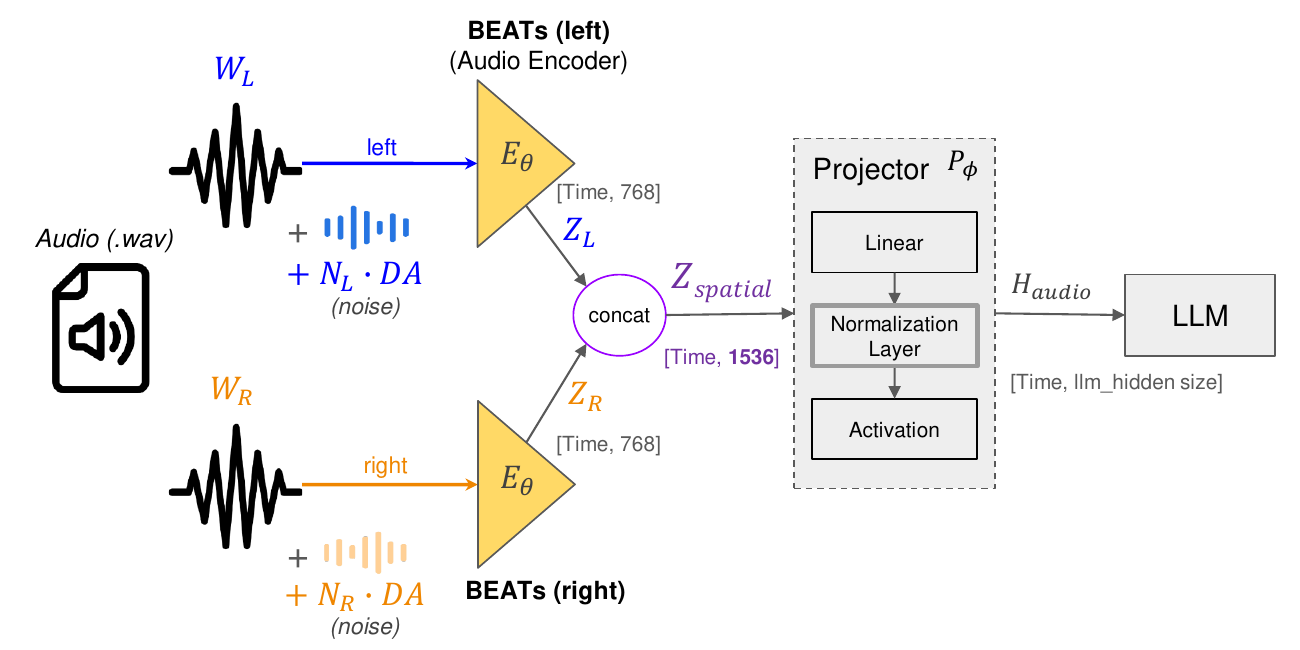}
    \caption{The Dual-BEATs Architecture. Uncorrelated dithering noise ($N_L, N_R$), scaled by a Dithering Amplitude ($DA$), is injected into decoupled stereo channels prior to encoding. The independent encodings are feature-concatenated ($Z_{spatial}$) before LLM ingestion.}
    \label{fig:dualbeats}
\end{figure}

To formalize the spatial routing topology illustrated in Figure~\ref{fig:dualbeats}, let $E_{\theta}$ map inputs to a latent dimension $d$ (e.g., $d=768$). Parallel forward passes yield independent temporal sequences:
$$Z_L = E_{\theta}(W_L'), \quad Z_R = E_{\theta}(W_R') \quad \in \mathbb{R}^{T \times d}$$

To construct spatial geometry, we fuse these representations along the feature dimension. Unlike sequence-dimension concatenation ($2T \times d$), which presents the channels as temporally disjoint events, feature-dimension concatenation enforces a strict temporal lock. This yields a unified, time-aligned payload $Z_{spatial} = [Z_L; Z_R] \in \mathbb{R}^{T \times 2d}$.

This 1536-dimensional representation passes through a learnable projector $P_{\phi}$, which utilizes activation and normalization functions mirroring the target LLM (e.g., LayerNorm \citep{ba2016layer} and SiLU \citep{elfwing2018sigmoid} for OLMo, RMSNorm \citep{zhang2019root} and GELU \citep{hendrycks2016gaussian} for Gemma) to map the acoustic features into the language model's latent space: $H_{audio} = P_{\phi}(Z_{spatial})$.

By projecting this time-aligned, feature-fused payload, Dual-BEATs transfers spatial localization directly to the LLM's native attention mechanisms. Because the left and right acoustic tokens are permanently bound within a single temporal step, the model is forced to dynamically evaluate the inter-channel amplitude delta—the literal volume and feature differences—to successfully reason about the physical acoustic space.

\subsection{The Normalization Bottleneck and Stochastic Resonance}
\label{normalization_bottleneck}

While Dual-BEATs routing successfully isolates the acoustic streams, empirical evaluations reveal a critical vulnerability within standard LLM topologies: internal variance normalization layers (e.g., RMSNorm, LayerNorm). Physical spatial geometry is encoded via subtle inter-channel amplitude differentials. At extreme spatial boundaries ($PA = 0.00$), this differential is large enough to survive regularization. However, as the sound source moves toward the center ($PA = 0.50$), the left and right acoustic features are separated only by microscopic variance. Standard normalizers act as aggressive compressors on these subtle continuous gradients, dynamically scaling the representations such that $\text{Norm}(Z_L) \approx \text{Norm}(Z_R)$. This effectively equalizes the channels, completely erasing the physical spatial geometry before the attention layers can process it.

\paragraph{The Uncorrelated Dithering Intervention}
To circumvent this compressive architectural bottleneck, we introduce a signal-level intervention inspired by Stochastic Resonance. Prior to encoding, we inject static, uncorrelated Gaussian noise directly into the continuous waveforms $W_L$ and $W_R$. Using a specific Dithering Amplitude ($DA$) scaling factor, we generate two strictly independent noise vectors from a standard normal distribution, $N_L, N_R \sim \mathcal{N}(0, 1)$, yielding the augmented signals:
$$W_L' = W_L + (N_L \cdot DA), \quad W_R' = W_R + (N_R \cdot DA)$$

Crucially, injecting strictly independent noise ($N_L \neq N_R$) establishes an artificial macro-variance floor. When these representations reach the LLM, the normalizer locks onto this dominant noise variance, safely floating the delicate acoustic amplitude delta across the architectural bottleneck. Furthermore, enforcing this uncorrelated property is essential to prevent shortcut learning \citep{geirhos2020shortcut}. If a shared noise vector were applied to center-panned audio ($W_L = W_R$), it would yield perfectly mirrored latent representations, allowing the model to achieve artificially inflated localization accuracy via a shallow identity comparison. Because real-world acoustic environments inherently contain microscopic phase and pressure discrepancies, enforcing uncorrelated noise deliberately breaks this artificial mathematical symmetry. The attention mechanisms are therefore forced to look past the dither and extract the true uncompressed spatial geometry, ensuring robust, real-world generalization rather than the exploitation of mathematical artifacts.

For all primary evaluations utilizing the Stochastic Resonance bridge, we establish the hyperparameter at $DA = 0.05$. This specific amplitude—approximating a -26 dB noise injection—was selected as a robust initial baseline. It ensures the injected macro-variance is substantial enough to survive the aggressive normalization compression without overwhelmingly obscuring the core semantic acoustic signal.

\subsection{Dataset Formulation and Spatial Instruction Tuning}

\paragraph{Dataset and Unified Task}
To evaluate our architecture, we utilize the official balanced splits of the AudioSet corpus \citep{gemmeke2017audio}, curated via the filtering pipeline established by Spatial-AST \citep{peng2024spatialast}. This yields 18,373 samples for training and 17,148 samples for evaluation across 355 distinct acoustic labels (see Appendix A.1 for exact reproducibility protocols and label distributions). Rather than separating spatial and semantic reasoning, we formulate a unified instruction objective: given a stereo input, the model must generate semantic event labels and physical spatial direction within a single continuous sequence.

\paragraph{Spatial Instruction Tuning and LoRA Adaptation}
To bridge the modality gap, we must align the randomly initialized projector $P_{\phi}$ and adapt the LLM to process the dithered stereo tokens. Crucially, empirical testing revealed that maintaining a strictly frozen LLM backbone resulted in a complete failure to learn spatial geometry. While the injected dither successfully preserves the inter-channel amplitude differentials across the normalization bottleneck, the LLM's native attention mechanisms are not naturally calibrated to correlate these microscopic variances. 

Therefore, we conduct a unified spatial instruction tuning phase. We maintain the dual-BEATs audio encoders in a strictly frozen state to preserve their robust semantic representations. To overcome the LLM's inherent spatial agnosticism while managing the massive memory footprint of uncompressed spatial tokens, we employ Quantized Low-Rank Adaptation (QLoRA) \citep{dettmers2023qlora}. We freeze the LLM backbone in a memory-efficient quantized state and apply higher-precision adapter modules directly to the core attention projections \citep{hu2022lora}. This allows the attention heads to adapt to the spatial geometry without inducing catastrophic forgetting of their semantic priors (exact quantization and LoRA configurations are detailed in Appendix B).

During this tuning phase, we expose the model to a 50/50 stochastic mixture of ``List-First'' and ``Direction-First'' instruction templates (detailed in Appendix A.2). This balanced prompt mixture forces the adapted attention heads to independently parse semantic and spatial features regardless of generation order, validating that our evaluations reflect true acoustic perception rather than autoregressive overfitting.

\paragraph{Strict Noise Floor Isolation}
To prevent the LLM from shortcut learning by memorizing the injected pseudo-random noise profiles, we enforce absolute stochastic isolation between phases. By strictly decoupling the random seeds used for noise generation during training and evaluation, we ensure all test-time noise signatures are fundamentally unseen. This guarantees that our localization accuracy reflects genuine inter-channel amplitude correlation rather than static artifact memorization (see Appendix A.3 for exact seed isolation protocols).

\section{Experiments and Results}
\label{experiments}

\subsection{Spatial Resolution and the Center-Pan Crash}

To evaluate the fundamental spatial capacity of the Dual-BEATs architecture, we analyzed directional accuracy (Left, Center, Right) across a spectrum of panning amplitudes ($PA$). We evaluated both architectures using a single-pass inference on the evaluation set. The results, detailed in Table~\ref{tab:main_results}, provide striking empirical evidence of the normalization bottleneck predicted in Section~\ref{normalization_bottleneck}, alongside the efficacy of the Stochastic Resonance bridge. 

\begin{table}[htb]
  \caption{Spatial directional accuracy (\%) across panning amplitudes ($PA$). Undithered baselines (Off) suffer catastrophic spatial collapse at subtle amplitudes ($PA = 0.50$). Conversely, activating the uncorrelated noise floor (On, $DA=0.05$) preserves continuous, high-resolution spatial perception.}
  \label{tab:main_results}
  \centering
  \begin{tabular}{l c c c c c c c c}
    \toprule
    & \multicolumn{2}{c}{$PA=0.00$} & \multicolumn{2}{c}{$PA=0.10$} & \multicolumn{2}{c}{$PA=0.25$} & \multicolumn{2}{c}{$PA=0.50$} \\
    \cmidrule(lr){2-3} \cmidrule(lr){4-5} \cmidrule(lr){6-7} \cmidrule(lr){8-9}
    Architecture & Off & On & Off & On & Off & On & Off & On \\
    \midrule
    \textbf{Gemma-3-1B + Dual-BEATs} \\
    $\rightarrow$ Direction-First & \textbf{66.8} & 47.7 & \textbf{80.8} & 80.4 & 36.7 & \textbf{70.2} & 33.8 & \textbf{58.0} \\
    $\rightarrow$ List-First & \textbf{61.4} & 53.0 & \textbf{91.6} & 68.1 & 37.0 & \textbf{58.0} & 33.2 & \textbf{45.6} \\
    \addlinespace
    \textbf{OLMo-3-7B + Dual-BEATs}  \\
    $\rightarrow$ Direction-First & \textbf{99.5} & 99.0 & 94.3 & \textbf{98.4} & 80.2 & \textbf{98.6}& 37.9 & \textbf{97.1} \\
    $\rightarrow$ List-First & \textbf{99.8} & 99.1 & 96.1 & \textbf{98.7} & 83.4 & \textbf{98.4} & 37.9 & \textbf{96.3} \\
    \bottomrule
  \end{tabular}
\end{table}

\paragraph{The Center-Pan Crash (Undithered Baseline)}
When evaluated without the dithering intervention, both architectures exhibit an extreme, non-linear degradation in spatial perception. At the binary acoustic extreme ($PA=0.00$)—where the inter-channel differential is large enough to survive layer normalization—the undithered OLMo-3-7B model effortlessly routes the signal, achieving 99.8\% accuracy. However, as the spatial gradient narrows toward the center ($PA=0.50$), the normalization compression triggers a catastrophic spatial collapse. The undithered OLMo-3-7B model's accuracy plummets to 37.9\%, while the Gemma-3-1B model falls to 33.2\%. Because random chance in this ternary task is 33.3\%, these scores indicate that the undithered models have been rendered completely spatially blind by the internal architectural bottleneck.

\paragraph{Stochastic Resonance Rescue}
Activating the uncorrelated dithering noise floor (DA 0.05) successfully circumvents this compressive bottleneck. By insulating the continuous inter-channel variance, the dithered OLMo-3-7B maintains a near-perfect, flat resolution curve across the entire panning spectrum, securing 97.1\% accuracy even at the most difficult $PA=0.50$ threshold. 

While the smaller Gemma-3-1B model demonstrates sensitivity to the noise-to-signal ratio at extreme amplitudes ($PA=0.00$)—where the dither briefly acts as an obfuscant rather than a carrier—it still strongly benefits from the resonance bridge at subtle amplitudes, rescuing it from random guessing (33.8\% $\rightarrow$ 58.0\%). However, the OLMo-3-7B results demonstrate the true ceiling of the Dual-BEATs architecture when supported by sufficient linguistic parameters.

\paragraph{Global Spatial Attention and Prompt Independence}
Crucially, the data demonstrates that this physical rescue translates to robust, generalized acoustic perception regardless of autoregressive load. Notably, the ``Direction-First'' formulation poses a significantly harder cognitive challenge, as the model must output the spatial localization token immediately, prior to unfolding the semantic sequence. Despite this, the dithered OLMo-3-7B model achieves its highest performance (97.1\%) in this exact mode. This confirms that the Dual-BEATs architecture successfully grants the model ``global'' spatial attention across the entire audio context window before token generation begins, definitively proving that standard LLMs are capable of generalized stereo perception without dedicated stereo audio encoders.

\subsection{Zero-Shot Spatial Generalization}
\begin{figure}[ht]
  \centering
  
  \begin{subfigure}{0.48\textwidth}
    \centering
    \includegraphics[width=\linewidth]{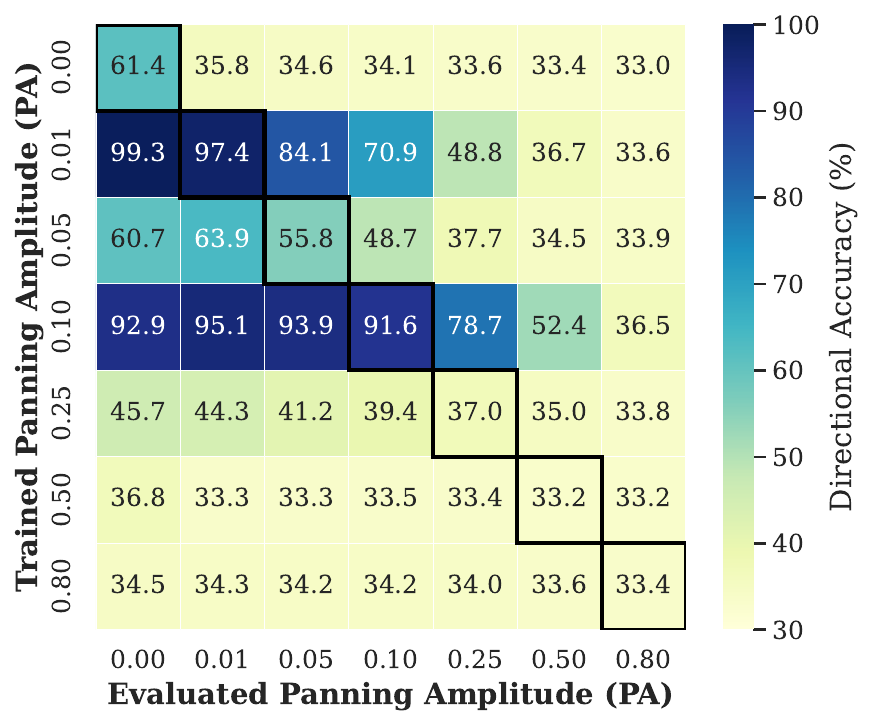}
    \caption{Gemma-3-1B-it (Undithered)}
    \label{fig:gemma_undithered}
  \end{subfigure}
  \hfill
  \begin{subfigure}{0.48\textwidth}
    \centering
    \includegraphics[width=\linewidth]{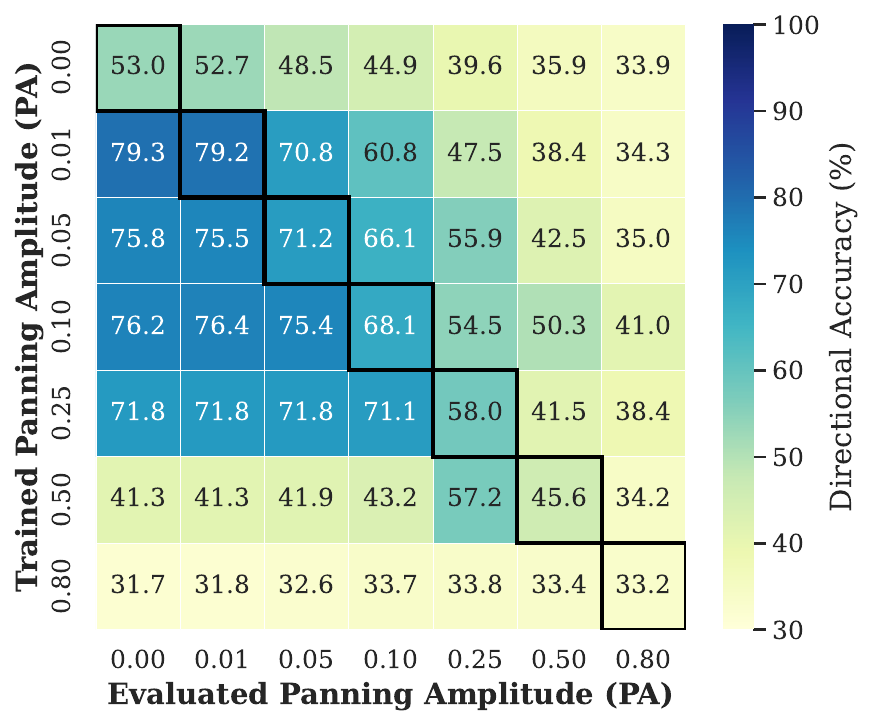}
    \caption{Gemma-3-1B-it (Dithered)}
    \label{fig:gemma_dithered}
  \end{subfigure}
  
  \vspace{0.5cm} 
  
  \begin{subfigure}{0.48\textwidth}
    \centering
    \includegraphics[width=\linewidth]{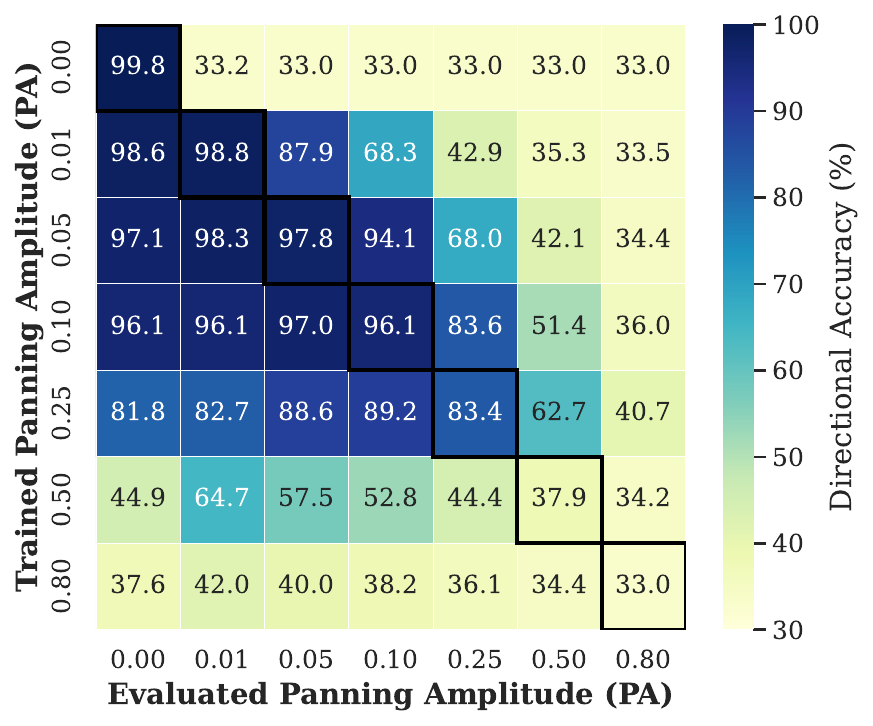}
    \caption{OLMo-3-7B (Undithered)}
    \label{fig:olmo_undithered}
  \end{subfigure}
  \hfill
  \begin{subfigure}{0.48\textwidth}
    \centering
    \includegraphics[width=\linewidth]{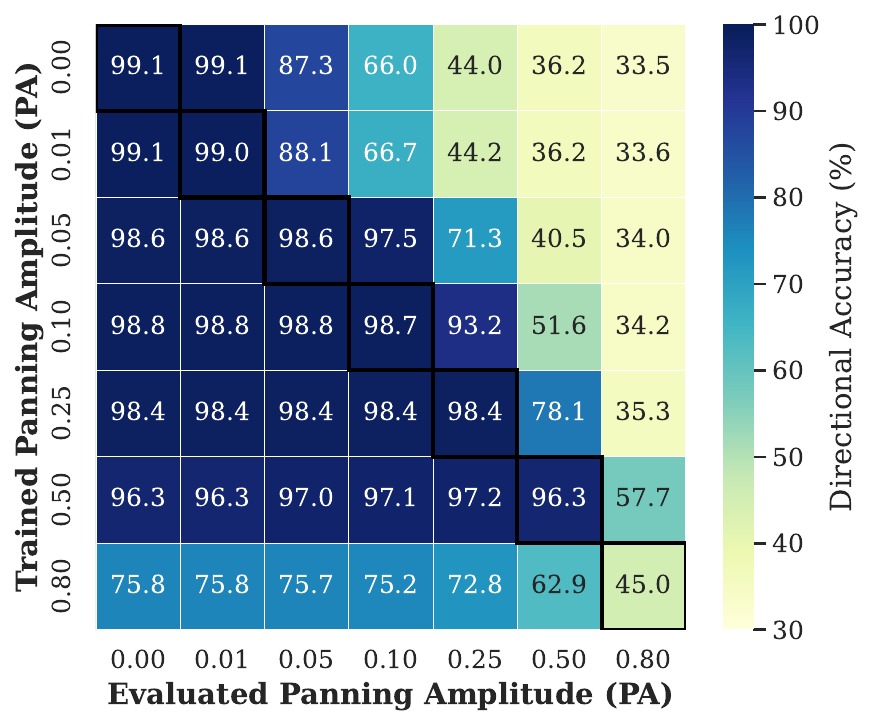}
    \caption{OLMo-3-7B (Dithered)}
    \label{fig:olmo_dithered}
  \end{subfigure}
  
  \caption{Zero-shot spatial generalization heatmaps (List-First condition). Undithered baselines (left) overfit to specific training amplitudes, exhibiting brittle off-diagonal fragmentation. In contrast, dithered models (right) achieve continuous, robust off-diagonal spatial mapping across entirely unseen acoustic geometries, validating the Stochastic Resonance bridge.}
  \label{fig:zero_shot_heatmaps}
  
\end{figure}

To verify that the Dual-BEATs architecture achieves true spatial abstraction rather than merely memorizing discrete training artifacts, we evaluated its zero-shot generalization capabilities. Models fine-tuned exclusively on a single panning amplitude (Y-axis) were evaluated across a continuous spectrum of entirely unseen spatial environments ranging from $PA=0.00$ to $PA=0.80$ (X-axis). The resulting performance matrices are visualized in Figure~\ref{fig:zero_shot_heatmaps}.

\paragraph{The Brittle Diagonal (Undithered Baseline)}
As demonstrated in the left column of Figure~\ref{fig:zero_shot_heatmaps}, the undithered baselines fundamentally fail to generalize. While they occasionally memorize their specific training amplitudes, off-diagonal performance collapses to the random-chance floor (~33.3\%). This brittleness is most devastatingly illustrated in the undithered OLMo-3-7B model trained at $PA=0.00$ (Figure~\ref{fig:olmo_undithered}, top row). While it achieves 99.8\% accuracy when evaluating identical $PA=0.00$ inputs, shifting the evaluated audio to an acoustically adjacent, nearly imperceptible $PA=0.01$ triggers an immediate collapse to 33.2\%. This confirms that without the dithering intervention, the LLM fails to learn continuous acoustic geometry. Instead, it overfits to extreme mathematical artifacts—specifically the literal zero-value tensors produced by a fully muted audio channel at $PA=0.00$.

\paragraph{Continuous Spatial Abstraction (Dithered Models)}
Conversely, activating the Stochastic Resonance bridge (right column) fundamentally transforms this behavior, replacing the fragmented diagonal with a massive, robust block of high-resolution spatial mapping. The OLMo-3-7B model demonstrates exceptional zero-shot capability across the spectrum. For example, a model trained exclusively on a mid-range amplitude ($PA=0.25$) successfully extrapolates to entirely unseen extreme boundaries ($PA=0.00$) with negligible degradation (98.4\%), while maintaining a robust 78.1\% on subtler panning amplitudes ($PA=0.50$). 

Even more remarkably, training the dithered OLMo-3-7B model on the highly compressed $PA=0.50$ setting (Figure~\ref{fig:olmo_dithered}, second row from bottom) acts as a universal spatial anchor. By forcing the model to resolve spatiality at this difficult acoustic threshold, it achieves over 96.3\% zero-shot accuracy across every single higher-differential amplitude ($PA=0.00$ to $PA=0.25$). This visual evidence solidifies the core architectural hypothesis: by safely floating the delicate spatial geometry across the normalization layers, the LLM is forced to learn the continuous physical gradient of the acoustic space, completely circumventing the need for domain-specific spatial encoders.

\paragraph{The Psychoacoustic Resolution Limit ($PA=0.80$)}
At the extreme boundary of subtlety ($PA=0.80$), the inter-channel differential is approximately 1.9 dB. This borders on the Just Noticeable Difference (JND) threshold within human psychoacoustics, causing models trained exclusively at this amplitude to struggle with spatial convergence (visible in the bottom rows of Figure~\ref{fig:zero_shot_heatmaps}). However, the robust spatial anchor formed by training the dithered OLMo-3-7B model at $PA=0.50$ allows it to extrapolate into this difficult threshold. As shown in Figure~\ref{fig:olmo_dithered}, it achieves a non-trivial 57.7\% zero-shot accuracy on $PA=0.80$ evaluations—significantly outperforming the random chance baseline even at the absolute edge of the architecture's physical resolution limit.

\subsection{Semantic Retention and the Semantic Tax}
\begin{figure}[t]
  \centering
  \includegraphics[width=\textwidth]{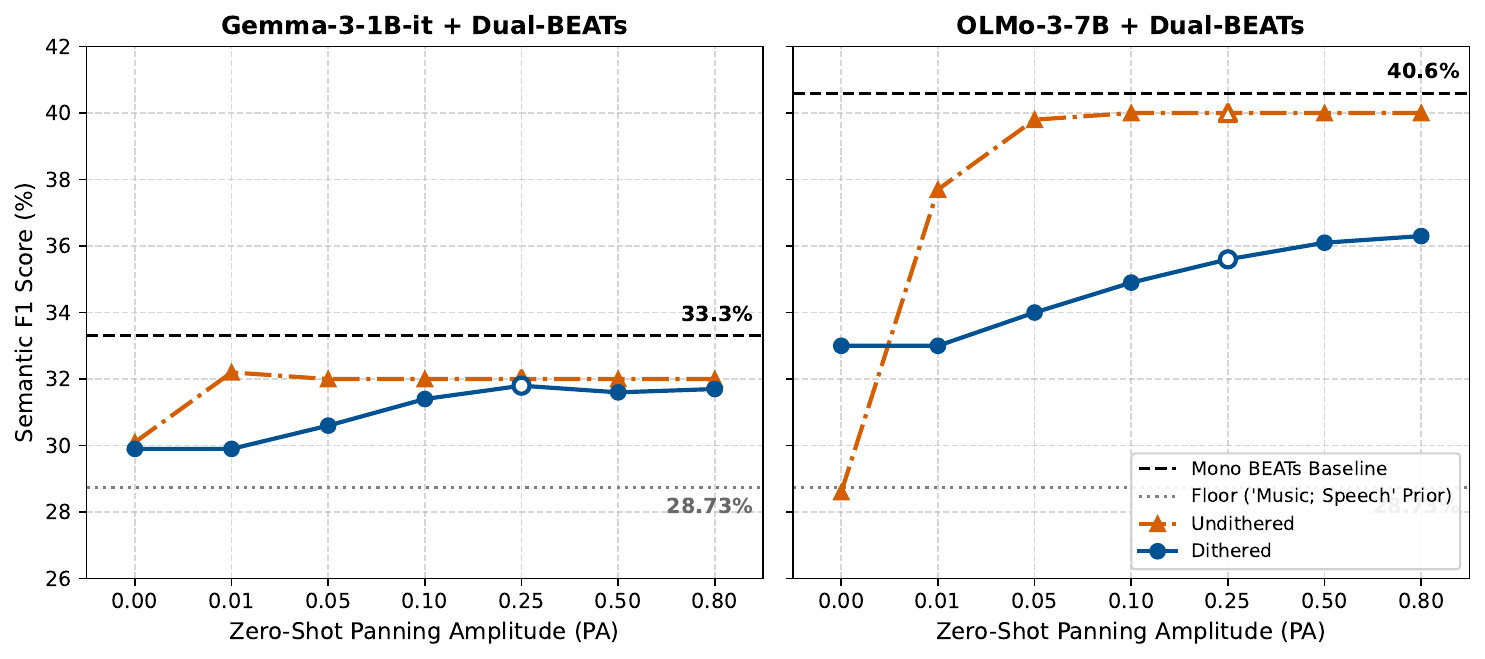}
  \caption{Zero-shot semantic retention evaluated under the strictly isolated ``List-First'' condition for models trained at $PA = 0.25$. Solid markers indicate zero-shot evaluations; hollow markers denote in-domain baselines. At the binary boundary ($PA = 0.00$), the undithered baseline (orange) suffers a normalization crash, reverting entirely to the 28.73\% statistical prior. The dithered architecture (blue) preserves stable semantic extraction across the entire spatial spectrum.}
  \label{fig:semantic_retention}
\end{figure}

While the Stochastic Resonance bridge flawlessly preserves spatial geometry, injecting static, uncalibrated noise inherently acts as an acoustic perturbation. To rigorously evaluate this trade-off, we isolated the semantic extraction performance using the ``List-First'' prompt to measure the resulting ``Semantic Tax.''

To properly contextualize these metrics, we must establish the dataset's statistical baseline. As detailed in Appendix A.1, the inherent class imbalance of the official AudioSet evaluation corpus creates a massive statistical prior. A naive, deterministic model that completely ignores the audio and exclusively guesses the two most naturally frequent labels (``Music'' and ``Speech'') achieves a semantic F1 score of 28.73\% without any genuine acoustic perception.

Figure~\ref{fig:semantic_retention} details the zero-shot semantic retention curve for a representative OLMo-3-7B model fine-tuned at a mid-range amplitude ($PA=0.25$). Under aggressive dithering ($DA=0.05$), the Dual-BEATs architecture experiences a stable, expected reduction in semantic capacity (shifting from the theoretical mono ceiling of 40.6\% to approximately 35.5\% F1). Crucially, this dithered performance remains completely flat and robust across all zero-shot spatial amplitudes.

In stark contrast, when the undithered baseline encounters the hard-panned $PA=0.00$ boundary, it violently collapses. The variance singularity mathematically shocks the normalizers, triggering a total acoustic ablation that drops the undithered model exactly down to the 28.73\% statistical floor. The noise floor's ability to safely float the semantic payload above this baseline crash definitively proves that continuous inter-channel geometry can be preserved without destroying the underlying acoustic tokens.

\section{Limitations and Broader Impacts}
\label{sec:limitations}

\paragraph{Acoustic and Environmental Limitations}
While the Dual-BEATs architecture successfully bypasses normalizer constraints to extract spatial geometry via Inter-channel Level Differences (ICLD), our current evaluations are bounded by single-source, amplitude-panned acoustic environments. The resolution of multi-source, overlapping audio from distinct spatial coordinates remains untested. Furthermore, we have not yet evaluated the Stochastic Resonance bridge against true binaural audio, which introduces highly complex, non-linear Head-Related Transfer Functions (HRTFs) and micro-second phase delays not captured by simple deterministic amplitude panning. Finally, our reliance on the AudioSet corpus introduces ontological biases—heavily skewing toward ``Speech'' and ``Music'' (see Appendix A.1)—meaning spatial acuity for underrepresented acoustic classes remains an open question for equitable deployment.

\paragraph{Architectural and Computational Constraints}
While uncorrelated dithering bypasses normalizer constraints, the static noise floor introduces a measurable ``Semantic Tax'' (an approximate 5\% reduction in baseline F1 score). Notably, our current evaluations used a fixed $DA = 0.05$. Because the normalization bottleneck equalizes microscopic variances aggressively—with undithered performance collapsing at a mere $PA = 0.01$-the minimum required variance floor may be lower than our baseline. Future work will map the dithering spectrum to optimize this trade-off and minimize semantic degradation. Additionally, our method requires an uncompressed feature density of 50 tokens per second. This sequence length, combined with FP32 upcasting to stabilize cross-modality gradients on hardware lacking native BF16 support, introduces substantial memory overhead during tuning. Finally, expanding this continuous spatial modality into larger language backbones (>10B parameters) risks severe modal shock, requiring datasets with massive volumetric density to stabilize projector gradients without inducing catastrophic unlearning.

\paragraph{Broader Societal Impacts}
The development of generalized spatial acoustic reasoning in Audio Large Language Models carries significant dual-use implications. On a positive societal vector, natively granting LLMs ``stereo hearing'' drastically expands the horizon for accessibility tools; for example, spatially aware AI assistants could provide critical environmental navigation cues (e.g., localizing sirens, hazards, or speakers) for visually impaired users. Conversely, the deployment of highly accurate, zero-shot acoustic scene analysis poses inherent privacy risks. If integrated into edge devices or public microphone arrays, this architecture could theoretically be misused for unauthorized, localized surveillance or the persistent acoustic tracking of individuals without their consent. Furthermore, the architecture’s reliance on a static noise floor introduces an adversarial vector: an actor intercepting the dither seed could inject imperceptible acoustic perturbations to spoof spatial perception. We urge the prioritization of robust consent frameworks and acoustic sanitization protocols as multimodal models transition toward physical environmental grounding.

\section{Conclusion}
This paper introduces the Dual-BEATs architecture, a novel multimodal routing topology that endows foundational Audio Large Language Models with true spatial acoustic perception without relying on specialized, geometry-aware spatial encoders. By physically decoupling the stereo signal, we empirically exposed a critical architectural vulnerability within standard LLM topologies: internal variance normalizers act as aggressive compressors on continuous audio, obliterating subtle inter-channel amplitude geometries and triggering catastrophic spatial collapse.

To overcome this bottleneck, we proposed a targeted, signal-level intervention inspired by Stochastic Resonance. By injecting an uncorrelated, static noise floor prior to semantic encoding, we established an artificial macro-variance that safely floats the spatial geometry across the normalization layers. Our evaluations demonstrate that this intervention not only preserves robust spatial resolution at highly compressed, center-panned boundaries, but it also unlocks profound zero-shot spatial extrapolation across entirely unseen acoustic environments. 

Ultimately, this work establishes that standard LLM cross-attention mechanisms are fundamentally capable of high-resolution continuous spatial reasoning. By properly shielding the physical geometry from standard regularization, we provide a highly modular, scalable foundation for the next generation of spatially-aware multimodal systems.

\ack{The successful completion of this research was made possible by the academic resources and advanced research infrastructure provided by the National Center for High-Performance Computing, National Institutes of Applied Research (NIAR), Taiwan. We gratefully acknowledge their invaluable support.}

{

\small
\bibliographystyle{unsrtnat}
\bibliography{references} 
}

\newpage

\appendix
\section{Appendix: Experimental Setup and Dataset Details}
\label{app:experiment}

\subsection{Dataset Curation and Label Distribution}
\label{appA:dataset}
Our evaluation uses the official balanced training and evaluation splits of the AudioSet corpus \citep{gemmeke2017audio}. To ensure direct comparability with specialized spatial baselines, we strictly follow the curation pipeline established by Spatial-AST \citep{peng2024spatialast}. Because AudioSet relies on live YouTube URLs, the live corpus naturally diminishes over time due to video deletions. To guarantee reproducibility against this link attrition, our dataset relies on a frozen third-party snapshot from March 2024. This snapshot preserves 18,373 samples (out of 22,160) for the training split and 17,148 samples (out of 20,371) for the evaluation split.

These available samples encompass a refined ontology of 355 distinct acoustic labels. However, the distribution is heavily skewed toward high-frequency priors: ``Music'' and ``Speech'' appear in 30.1\% and 28.5\% of evaluation samples, respectively (see Table~\ref{tab:label_dist}).

\begin{table}[ht]
\centering
\caption{Top 10 Most Frequent Acoustic Labels in the curated AudioSet subset.}
\label{tab:label_dist}
\small
\begin{tabular}{@{}llccc@{}}
\toprule
\textbf{Rank} & \textbf{Label} & \textbf{Train Freq. (\%)} & \textbf{Eval Freq. (\%)} & \textbf{Eval Count ($N$)} \\ \midrule
1 & Music & 31.20\% & 30.13\% & 5,166 \\
2 & Speech & 28.58\% & 28.54\% & 4,894 \\
3 & Vehicle & 4.49\% & 4.34\% & 744 \\
4 & Animal & 3.77\% & 3.78\% & 648 \\
5 & Musical instrument & 2.82\% & 2.29\% & 393 \\
6 & Singing & 2.63\% & 1.95\% & 334 \\
7 & Domestic animals & 2.13\% & 1.90\% & 325 \\
8 & Guitar & 1.86\% & 1.70\% & 292 \\
9 & Car\textsuperscript{\textdagger} & 1.55\% & 1.56\% & 267 \\
10 & Water\textsuperscript{\textdagger} & [---] & 1.53\% & 263 \\ \bottomrule
\end{tabular}
\vspace{0.2cm}

\footnotesize{\textsuperscript{\textdagger} \textit{Ranked by Eval frequency. In the Training set, ``Plucked string instrument'' appeared at Rank 9 (1.61\%, $N=295$) and ``Car'' appeared at Rank 10.}}
\end{table}

\paragraph{Defense Against Shortcut Learning}
The prevalence of these tags presents a theoretical opportunity for shortcut learning. However, our 97.2\% localization accuracy remains consistent across the 345 low-frequency categories. This confirms the Dual-BEATs architecture is performing genuine acoustic scene analysis rather than exploiting label co-occurrence.

\subsection{Unified Instruction Task and Target Schema}
\label{appA:instruction}
We format the raw audio and labels into a unified instruction objective, requiring the model to simultaneously extract semantic events and localize their spatial origin. To mitigate autoregressive bias based on sequence order, we apply a 50/50 stochastic split between two instruction variants:
\begin{itemize}
\item \textbf{List-First:} \textit{<audio> List the audio events, then state their spatial direction.''} \item \textbf{Direction-First:} \textit{ State the spatial direction, followed by the audio events.''}
\end{itemize}

While we adopt the semantic ontology of Spatial-AST \citep{peng2024spatialast}, we condense the output into a single conversational response, detailed in Table~\ref{tab:prompts}.

\begin{table}[ht]
\centering
\caption{Unified Prompt Text and Target Response Schema}
\label{tab:prompts}
\small
\begin{tabular}{@{}lp{5.5cm}p{5.5cm}@{}}
\toprule
\textbf{Mode} & \textbf{Input Prompt} & \textbf{Target Response Format} \\ \midrule
\textbf{List-First} & ``<audio> List the audio events, then state their spatial direction.'' & \texttt{[Event 1]; [Event 2]; ...; Direction: [d]} \\
\textbf{Direction-First} &``<audio> State the spatial direction, followed by the audio events.'' & \texttt{From the [d]: [Event 1]; [Event 2]; ...} \\ \bottomrule
\end{tabular}
\end{table}

\subsection{Amplitude Panning}
\label{appA:panning}
During data loading, we spatialize mono sources $W$ dynamically into stereo pairs $(W_L, W_R)$ using deterministic amplitude panning. Samples have an equal 33.3\% probability of Center, Left, or Right assignment, scaled by a Panning Amplitude ($PA$):
\begin{itemize}
    \item \textbf{Center:} $W_L = W, \quad W_R = W$
    \item \textbf{Left:} $W_L = W, \quad W_R = W \cdot PA$
    \item \textbf{Right:} $W_L = W \cdot PA, \quad W_R = W$
\end{itemize}

While computationally simpler than room-impulse simulations, this method cleanly isolates Inter-channel Level Differences (ICLD). Table~\ref{tab:db_diff} translates experimental $PA$ values into approximate acoustic decibel (dB) differentials.

\begin{table}[ht]
\centering
\caption{Inter-channel Level Differences (ICLD) by Panning Amplitude}
\label{tab:db_diff}
\small
\begin{tabular}{@{}lcc@{}}
\toprule
\textbf{Panning Amplitude ($PA$)} & \textbf{Attenuated Channel Multiplier} & \textbf{Approx. Acoustic Difference ($\Delta$ dB)} \\ \midrule
0.8 & $W \cdot 0.8$ & 1.9 dB \\
0.5 (Primary Eval) & $W \cdot 0.5$ & 6.0 dB \\
0.25 & $W \cdot 0.25$ & 12.0 dB \\
0.10 & $W \cdot 0.10$ & 20.0 dB \\
0.05 & $W \cdot 0.05$ & 26.0 dB \\
0.01 & $W \cdot 0.01$ & 40.0 dB \\ \bottomrule
\end{tabular}
\end{table}

\subsection{Empirical Analysis of Shortcut Learning and Dither Correlation}
\label{appA:dither}
Our strict requirements for uncorrelated noise and absolute seed isolation were driven by empirical observations of severe shortcut learning during early architectural ablations. Large Language Models are highly susceptible to exploiting superficial statistical artifacts to minimize loss, bypassing the intended acoustic scene analysis.

\paragraph{The Threat of Correlated Noise}
During initial testing with \textit{correlated} dithering noise ($N_L = N_R$), a spatially-deaf baseline unexpectedly achieved 66.7\% accuracy. Analysis revealed the LLM acted as a superficial comparator: since $W$ and $N$ were identical, Center-panned audio yielded mathematically identical Left and Right embeddings. The model learned a binary shortcut: predict ``Center'' if embeddings match ($Z_L = Z_R$), otherwise guess randomly between ``Left'' and ``Right''. 

This comparator shortcut can be eliminated by enforcing strictly \textit{uncorrelated} noise vectors ($N_L \neq N_R$) so that the Left and Right embeddings are never identical, thereby forcing the model to learn true Inter-channel Level Differences (ICLD).

\paragraph{Seed Memorization and Conflicting Cues}
We additionally observed severe performance deterioration when the random seed was not strictly isolated between training and evaluation phases ($S_{train} = S_{eval}$). Because the pseudo-random number generation stream is deterministic, reusing the training seed during evaluation generates the exact same sequence of uncorrelated noise tensors ($N_L, N_R$), albeit superimposed onto novel evaluation audio features. Through observation, we suspected that the LLM may be capable of memorizing these specific noise structures as superficial shortcuts for spatial direction.

Consequently, our methodology strictly mandates $S_{eval} \neq S_{train}$ to neutralize this shortcut. We enforce absolute random seed isolation by using $S=42$ for training and a completely unseen seed $S=1337$ for all evaluations. This isolation rigidly controls the prompt mixture selection, the spatial panning distribution, and the injection of the uncorrelated dithering noise floors ($N_L, N_R$), ensuring that our high localization accuracy reflects true zero-shot acoustic perception.

\section{Appendix: Implementation and Optimization Details}
\label{app:implementation}

\subsection{Hardware constraints and High-Resolution Token Density}
\label{appB:hardware}
Training utilized NVIDIA V100 GPUs (32GB VRAM). To preserve microscopic amplitude differentials from dithering, we bypassed temporal pooling, maintaining an uncompressed feature density of 50 tokens per second. While vital for localization, this significantly expands the memory footprint. To accommodate this within V100 constraints, we used gradient accumulation to achieve an effective global batch size of 16.

\subsection{Cross-Modality Numerical Stability and Projection}
Aligning modalities across differing native precisions (e.g., Gemma's BF16 vs. BEATs' continuous features) on hardware lacking native BF16 acceleration risks gradient underflow. To ensure stability, FP16 acoustic features were explicitly upcasted to FP32 during the forward pass. The linear projector $P_{\phi}$ matches the exact hidden dimensions and architectural stacks of the respective LLMs:
\begin{itemize}
    \item \textbf{Gemma-3-1B-it Mapping:} Projector output dimension of 1152, utilizing an RMSNorm + GELU activation stack.
    \item \textbf{OLMo-3-7B-Instruct Mapping:} Projector output dimension of 4096, utilizing a LayerNorm + SiLU activation stack.
\end{itemize}

\subsection{Optimization and LoRA Adaptation Configuration}
\label{appB:lora}
Optimization used AdamW with a cosine scheduler, a 0.1 warmup ratio, a 0.3 max gradient norm, and ran for one epoch.

\paragraph{Base Model Quantization (QLoRA)}
To fit the uncompressed context window on 32GB GPUs, base LLMs were frozen in 4-bit NormalFloat (NF4) using QLoRA \citep{dettmers2023qlora}. Computations remained in FP32 to protect delicate spatial gradients, bridging the FP32 projection layer with FP16 adapters applied to the attention mechanisms. Training hyperparameters and target modules are detailed in Table~\ref{tab:hyperparams}.

\begin{table}[ht]
\centering
\caption{Fine-Tuning Hyperparameters and LoRA Target Modules}
\label{tab:hyperparams}
\small
\begin{tabular}{@{}lcc@{}}
\toprule
\textbf{Hyperparameter} & \textbf{Gemma-3-1B-it} & \textbf{OLMo-3-7B-Instruct} \\ \midrule
Optimizer & AdamW & AdamW \\
Peak Learning Rate & 2e-5 & 1e-5 \\
LR Scheduler & Cosine & Cosine \\
Warmup Ratio & 0.1 & 0.1 \\
Max Gradient Norm & 0.3 & 0.3 \\
Effective Batch Size & 16 & 16 \\
Training Epochs & 1 & 1 \\
Base Backbone Precision & 4-bit (NF4) & 4-bit (NF4) \\
Compute / Adapter Precision & FP16 (FP32 Bridge) & FP16 (FP32 Bridge)\\
\midrule
\textbf{LoRA Target Modules} & \texttt{q\_proj, k\_proj, v\_proj, o\_proj,} & \texttt{qkv\_proj, o\_proj,} \\
 & \texttt{gate\_proj, up\_proj, down\_proj} & \texttt{gate\_up\_proj, down\_proj} \\
\textbf{Modules to Save} & \texttt{embed\_tokens, lm\_head} & \texttt{embed\_tokens, lm\_head} \\ \bottomrule
\end{tabular}
\end{table}

\subsection{Evaluation Inference Hardware}
Zero-shot evaluations were executed on NVIDIA H100 GPUs (80GB VRAM) with a batch size of 64 in native bfloat16 (BF16). We enforced greedy decoding to ensure deterministic metric extraction. Mapping the full zero-shot spatial matrix (Figure~\ref{fig:zero_shot_heatmaps}) required 196 full-dataset evaluations (14 inference passes across 14 trained models), demanding approximately 82 total GPU hours of compute.


\end{document}